\documentclass[a4paper]{article}

\usepackage{INTERSPEECH2018}

\title{R-CRNN: Region-based Convolutional Recurrent Neural Network \\for Audio Event Detection}
\name{Chieh-Chi Kao, Weiran Wang, Ming Sun, Chao Wang}
\address{
  Amazon Alexa}
\email{\{chiehchi,weiranw,mingsun,wngcha\}@amazon.com}

\begin{document}

\maketitle
\begin{abstract}
This paper proposes a Region-based Convolutional Recurrent Neural Network (R-CRNN) for audio event detection (AED).
The proposed network is inspired by Faster-RCNN~\cite{FRCNN}, a well-known region-based convolutional network framework for visual object detection.
Different from the original Faster-RCNN, a recurrent layer is added on top of the convolutional network to capture the long-term temporal context from the extracted high-level features.
While most of the previous works on AED generate predictions at frame level first, and then use post-processing to predict the onset/offset timestamps of events from a probability sequence; the proposed method generates predictions at event level directly and can be trained end-to-end with a multi-task loss, which optimizes the classification and localization of audio events simultaneously.
The proposed method is tested on DCASE 2017 Challenge dataset~\cite{DCASE2017challenge}.
To the best of our knowledge, R-CRNN is the best performing single-model method among all methods without using ensembles both on development and evaluation sets.
Compared to the other region-based network for AED (R-FCN~\cite{Kaiwu2017}) with an event-based error rate (ER) of 0.18 on the development set, our method reduced the ER to half.
\end{abstract}

\section{Introduction}
Audio event detection (AED) aims to enable intelligent systems to understand the surrounding environment based on audio cues. 
AED is well-suited for certain scenarios when other indicators (e.g. visual data) is not suitable or not available. 
AED has been appiled to autonomous driving and driving assistance systems to prevent accidents, where visual detection of the event is difficult when the object is not in sight.
Dobre et. al~\cite{SirenDetection} used AED to detect sirens on the road, and Foggia et. al~\cite{AED_AutonomousDriving} applied it to identify hazardous situations such as tire skidding and car crashes.
AED is also suitable for the surveillance in public transportation, where the visual analysis is not sufficient to reliably understand passengers activity due to the occlusions in overcrowded environments. 
Rouas et al.~\cite{AED_PublicTransportation} used AED to detect critical situations and to warn the control room, and Laffitte et al.~\cite{AED_Screaming} applied AED to detect scream and shouted speech in subway trains.
It is a common challenge to collect data for training a robust detector when these target events only happen occasionally in the real-world scenario.
The Detection and Classification of Acoustic Scenes and Events (DCASE) challenges have been held since 2013~\cite{DCASE2013,DCASE2016,DCASE2017challenge} to stimulate the research in this field.
DCASE Challenge 2017~\cite{DCASE2017challenge} task 2 provides datasets and baseline systems for detection of rare sound events, which asks to identify the onset time of target events (baby crying, glass breaking, and gunshot) within synthesized 30-second clips.

Among the submitted systems in DCASE 2017, most of the models consist of Deep Neural Network (DNN), Convolutional Neural Network (CNN), and Recurrent Neural Network (RNN).
These works make frame level prediction followed by post-processing to generate the hypothesis of audio events.
The baseline system~\cite{DCASE2017challenge} takes a chunk of spectrogram as input, and then feed it into one CNN and one RNN. 
Outputs from these two networks are fed it into a DNN for final classification.
The methods ranked top 2 in the challenge~\cite{Lim2017,Cakir2017} apply Convolutional Recurrent Neural Network as the main architecture.
Lim et al.~\cite{Lim2017} used 1D CNN with 2 layers of long short term memory (LSTM) layers to generate the frame level prediction; 
Cakir et al.~\cite{Cakir2017} used 2D CNN with gated recurrent unit (GRU) layers to compute the prediction at each frame.
These frame level predictions need a post-processing stage to generate onset/offset timestamps for target events.
Median filters~\cite{Phan2017,Cakir2017} and ad-hoc rules~\cite{Lim2017} have been used as the post-processing step. 
In this work, we make predictions at event level, and onset/offset timestamps are incorporated into the cost function directly.
There is no post-processing for converting frame level predictions (frame-wise probability) to event level predictions (onset/offset timestamps with event probability) for R-CRNN.

To make predicions at event level, region-based neural network is used as the main architecture.
Our model is inspired by Faster-RCNN~\cite{FRCNN}, a well known region-based convolutional network framework for visual object detection.
Different from detecting visual objects from an image, which has no temporal information included, AED can take advantage of temporal contextual information from a spectrogram.
Therefore, a recurrent layer is added on top of the convolutional network to capture the long-term temporal context from the extracted high-level features.
By incorporating temporal context, R-CRNN has better performance than the other region-based network for AED (R-FCN)~\cite{Kaiwu2017}, which is a fully convolutional network without recurrent layers.

\begin{figure}[t]
  \begin{center}
  \includegraphics[width=0.4\textwidth]{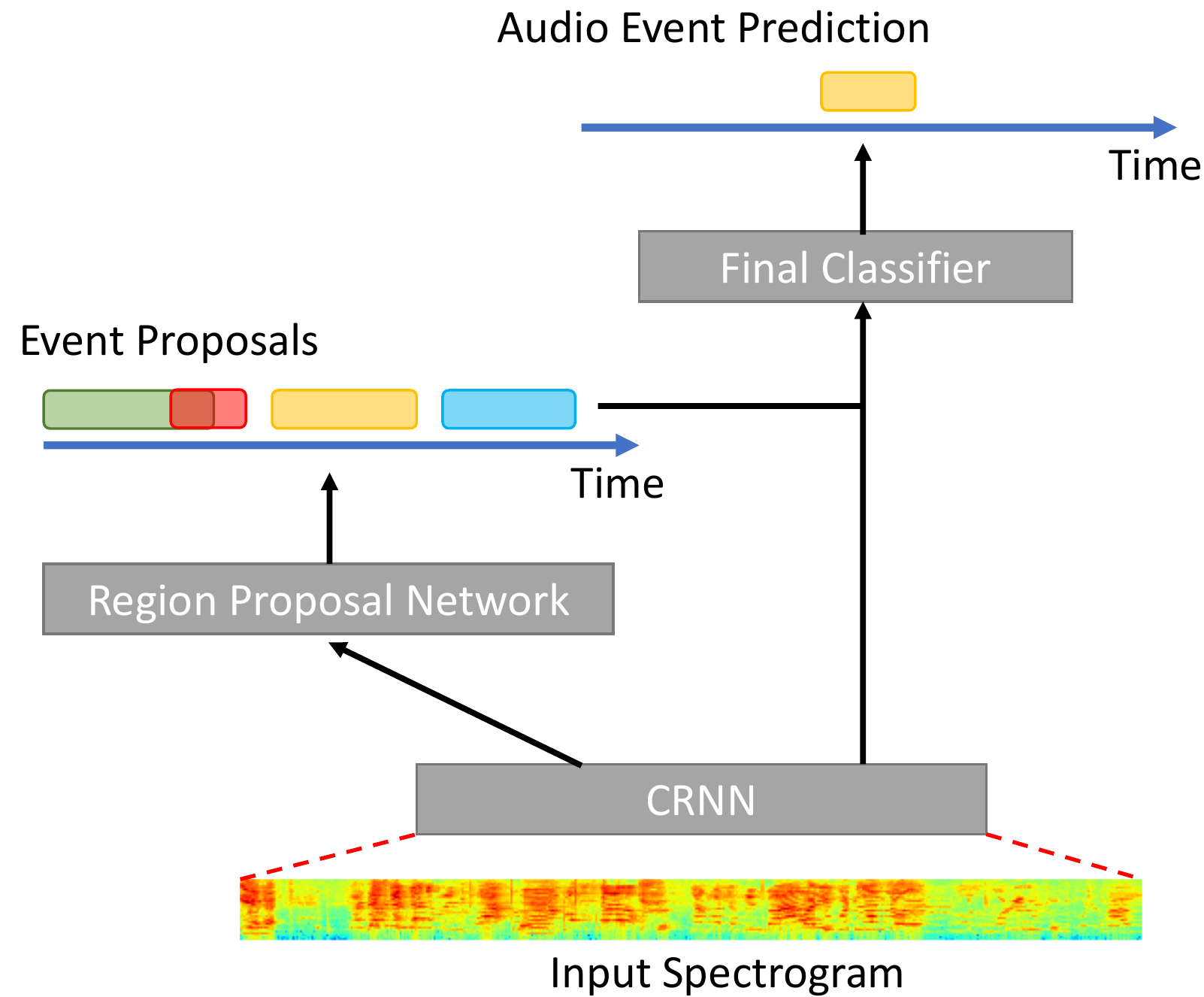}
  \caption{R-CRNN is an end-to-end model that predicts audio events at event level. No post-processing is needed for converting frame level to event level predictions.}
  \label{R-CRNN}
  \end{center}
\end{figure}

\section{R-CRNN}
As shown in Figure~\ref{R-CRNN}, the proposed R-CRNN consists of three modules: CRNN, region proposal network (RPN), and a final classifier.
The first CRNN module extracts high-level features from the input spectrogram.
The RPN computes event proposals based on the extracted features, and the classifier further refines the center/length of event proposals to generate audio event prediction.

\begin{figure}[t]
  \begin{center}
  \includegraphics[width=0.34\textwidth]{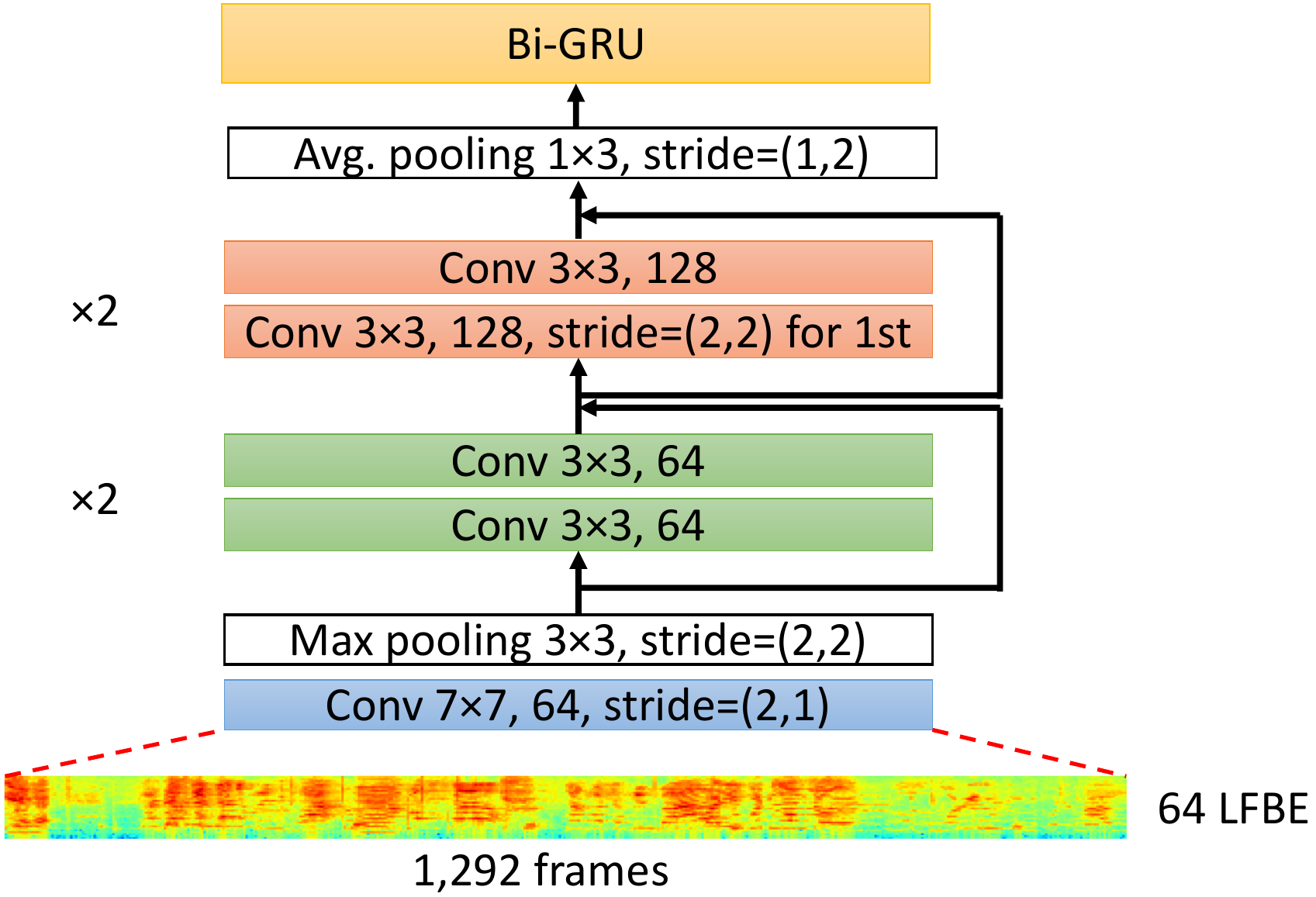}
  \caption{The architecture of CRNN. We use CRNN as a feature extractor in R-CRNN for AED. For the notation of convolutional kernels and strides, the first and second dimensions represent the time axis and frequency axis repectively.}
  \label{CRNN}
  \end{center}
\end{figure}

\subsection{Convolutional Recurrent Neural Network}
Figure~\ref{CRNN} shows the architecture of the proposed CRNN for feature extraction. 
CRNN takes 30-second sound clips as input and extracts the high-level feature map. 
We follow the settings in~\cite{Lim2017} to decompose each clip into a sequence of 46 ms frames (2,048 points sampled at 44.1k Hz) with a 23 ms shift.
64 dimensional log filter bank energies (LFBEs) are calculated for each frame, and we aggregate the LFBEs from all frames to generate the input spectrogram.
We use residual network (ResNet)~\cite{ResNet} as the convlutional network in CRNN, and there are two convolutional blocks in it.
2D convolutional kernels are used in CRNN, which generates a high-level feature map with time resolution of 186 ms (8$\times$ of the time resolution in the input spectrogram).
The size of the high-level feature map is (162, 2$U$), where $U$ is the number of units in the bi-directional GRU layer.

\begin{figure}[t]
  \begin{center}
  \includegraphics[width=0.38\textwidth]{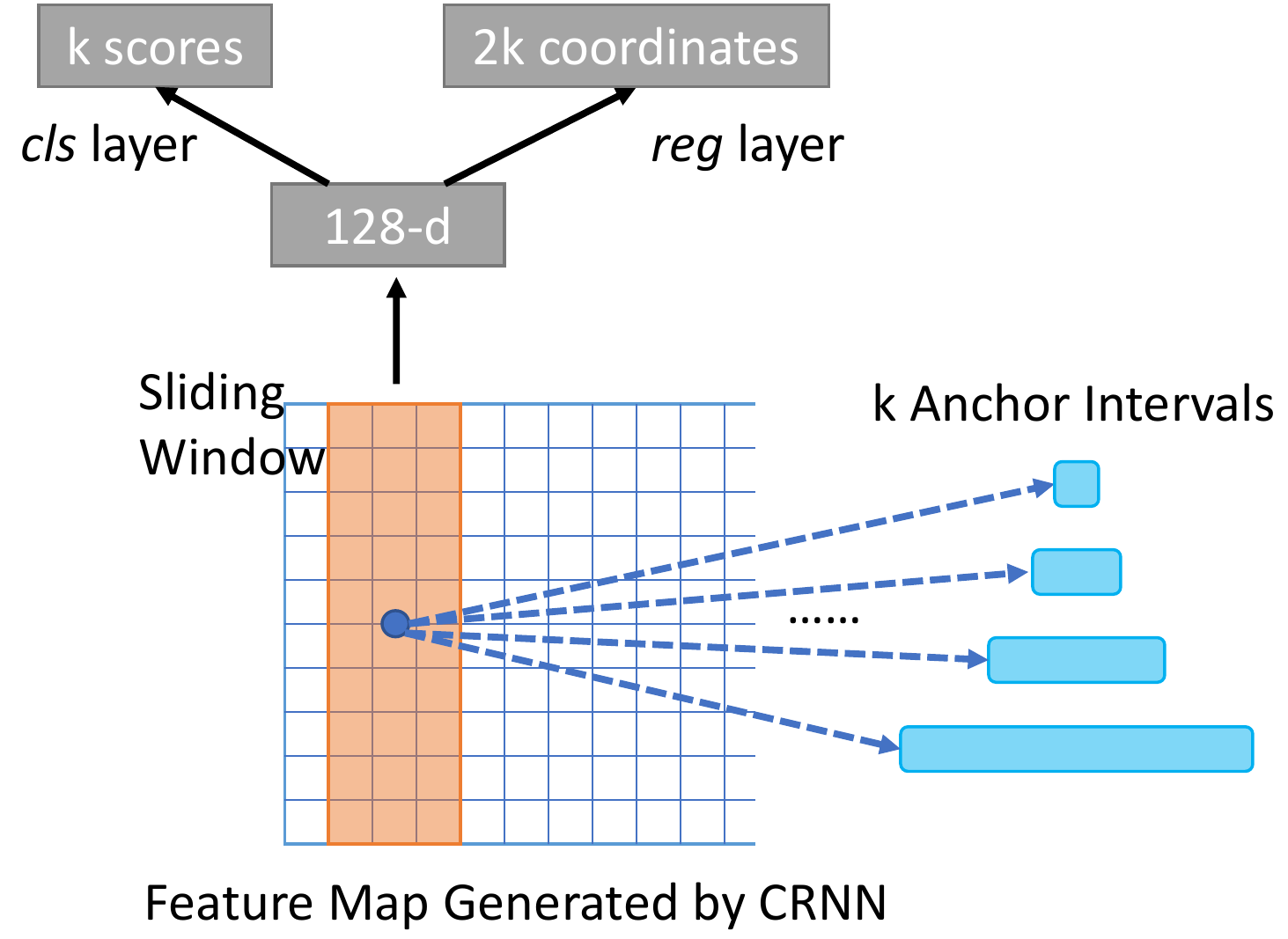}
  \caption{RPN generates event proposals by refining anchor intervals at each location on the time axis.}
  \label{RPN}
  \end{center}
\end{figure}

\subsection{Region Proposal Network}
We use RPN proposed by Ren et al.~\cite{FRCNN} to generate event proposals (time intervals) for AED.
While the original RPN generates region proposals in a 2-d search space ($x$, $y$ axes in an image), we simplify it to generate event proposals in a 1-D search space (time axis only).
In this work, RPN takes a region of the high-level feature map generated by CRNN, and outputs a set of event proposals, where each proposal has a probability of containing audio events.
We use RPN to quickly locate a vicinity of the event and reduce the number of intervals to be considered in the final classification.

As shown in Figure~\ref{RPN}, a stripe window slides over the high-level feature and maps the window to a lower-dimensional (128-d) feature.
The size of sliding window is $3{\times}n$, where $n$ is the height of the high-level feature map ($n=2U$ in our case). 
The receptive field of the sliding window is 557 ms ($3{\times}8{\times}$frame shift), but it can actually contain contextual information from intervals longer than 557 ms since the feature extractor CRNN has a recurrent layer.

At each frame of the high-level feature map, we propose multiple regions of different sizes center around it.
RPN takes anchor intervals with fixed sizes and then outputs $k$ interval proposals by refining these anchor intervals at each frame. 
The fixed sizes of anchor intervals are \{1,2,4,8,16,32\} frames in the high-level feature map through our experiments.
The 128-d feature of the sliding window at each location is fed to one dense layer ($cls$) to predict the probability of having an event ($k$ scores), and another dense layer ($reg$) to encode the coordinates of interval proposals ($2k$ coordinates).
Following the settings in~\cite{FRCNN}, these $k$ proposals are parameterized by shifting and scaling relative to $k$ anchor intervals.

For training RPN, each anchor interval is assigned to a ground-truth binary label that indicates containing target events or not.
Similar to the cost function defined in~\cite{FRCNN}, the cost function of RPN can be defined as:
\begin{equation}
L(\{p_i\},\{t_i\})=\sum_{i}{L_{cls}(p_i,p_i^*)}+\lambda\sum_{i}{p_i^*L_{reg}(t_i,t_i^*)},
\label{eq:RPN}
\end{equation}
where $i$ is the index of anchor interval, and $p_i$ is the predicted probability of containing target events for anchor $i$.
If anchor interval $i$ is highly overlapped with target events, the ground-truth label $p_i^*$ is set to one. 
If not, $p_i^*$ is set to zero.
$L_{cls}$ is the cross entropy for binary classification.
For the regression part, $t_i$ is a vector representing the two parameterized coordinates of the predicted interval proposal, and $t_i^*$ is the vector of the ground-truth event interval assigned to a positive anchor.
For $L_{reg}$, we use the robust loss function (smooth L1) defined in~\cite{FastRCNN}.
$\lambda$ is the coefficient to balance the classification error and regression error, and is set to one in all of our experiments.
This multi-task cost function (\ref{eq:RPN}) optimizes binary classfication and localization simultaneously.

\begin{figure}[t]
  \begin{center}
  \includegraphics[width=0.38\textwidth]{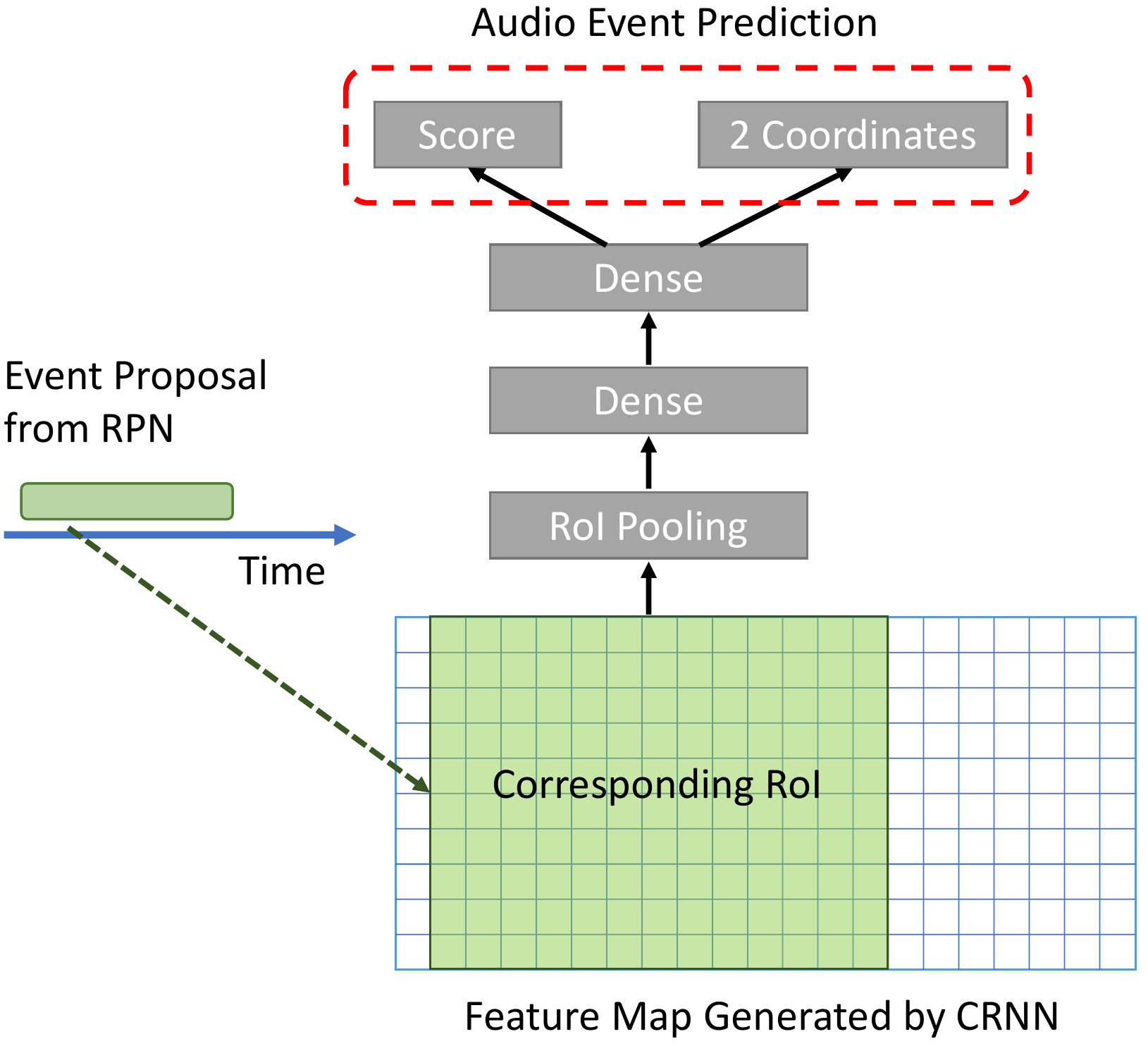}
  \caption{The final classifier takes event proposals generated from RPN and generates audio event predictions.}
  \label{classifier}
  \end{center}
\end{figure}

\subsection{Final Classifier}
After the interval proposals are available, we use non-maximum suppression (NMS)~\cite{NMS} to remove highly overlapped proposals.
In our experiments, 100 proposals based on their probability of containing audio events ($cls$) are selected to feed into the final classifier.
The final classifier takes event proposals generated from RPN as input, and generates audio event predictions.
As shown in Figure~\ref{classifier}, a corresponding region of interest (RoI) on the high-level feature map is cropped for each event proposal.
The cropped region is fed into a RoI pooling layer as proposed in ~\cite{FastRCNN} to generate a fixed-length feature vector ($7{\times}n$ in our experiments).
This fixed-length feature vector is fed into two dense layers ($M$ nodes) with dropout rate 0.5, followed by two output layers, which produce the probability of containing the target event, and the regression to refine the center/length of the event.
We use NMS again to remove highly overlapped events in these predictions.
Also, events with probability lower than a pre-set threshold ($0.8$) will be removed at this stage.
Since DCASE 2017 task 2 has an assumption that there is at most one event in each clip, only the event with the highest probability is kept in our experiments.

We use the same loss function (\ref{eq:RPN}) as RPN to train the final classifier, where the $i$ in becomes the index of event proposals.
This loss is used twice when training R-CRNN: one for RPN to identify the interval proposals, and one for the final classifier to predict audio events from event proposals.


\subsection{Extension for Multi-Class Event Detector}
R-CRNN can be extended to detect different classes of events simultaneously within the same framework by adding more outputs to the final classifier.
There is no change needed in RPN since it generates interval proposals agnostic to event classes.
For adding an extra class, the number of extra parameters is 3M+3 (M+1 for the classification and 2M+2 for the localization), where M is the number of nodes in the dense layers of the final classifier.

\section{Experimental Results}

\subsection{Dataset}
We tested R-CRNN on the dataset provided by DCASE 2017 Challenge task2: ``Detection of rare sound events.''
It consists of isolated target events (baby crying, glass breaking, and gunshot) downloaded from freesound.org, and 30-second background sound clips from TUT Acoustic Scenes 2016 dataset~\cite{TUT2016}.
A synthesizer provided in the development set is used to generate mixtures of target events and background sound clips at random onset time.
The synthesized clips are 30-second monaural audio with 44,100 Hz and 24 bits. 
For each target event, two training sets of 5,000 samples is generated with event-to-background ratios (EBR) of -6, 0, 6dB.
The first one is for the pre-training of CRNN, and the event occurrence probability is set to 0.5 for the synthesizer.
This event occurrence probability follows the same setting of development and test set provided by the challenge. 
The second one is for the training of R-CRNN, and the event occurrence probability is set to 0.99 to get more positive training samples.

\subsection{Pre-training}
From the experience of computer vision community, training a region-based network for detection from scratch is very difficult.
Previous works on visual object detection~\cite{FastRCNN,FRCNN} used weights pre-trained from image classification on ImageNet~\cite{ImageNet}  as the weight initialization for the feature extractor~\cite{VGG,ResNet} in region-based networks. 
In the pre-training, we pre-train the CRNN on a binary classification task using utterance level labels, and the trained CRNN is used as the weight initialization in the training of R-CRNN.
Note that given the pre-trained CRNN as the weight initialization, the training of R-CRNN is an end-to-end process, which optimizes the classification and localization of audio events simultaneously.
The pre-training stage is stopped when the loss on the development set has stopped improving for 10 epochs.
We use adaptive momentum (ADAM)~\cite{ADAM} as the optimizer and the initial learning rate is set to 0.001.
The size of mini-batch is set to 80.
We use Keras with Tensorflow backend to implement our models in the experiments.

\subsection{Training}
After the pre-training stage, the weights of pre-trained CRNN are used as the weight initialization of R-CRNN.
The training of R-CRNN is stopped when the loss on the development set has stopped improving for 10 epochs. 
We use ADAM as the optimizer and the initial learning rate is set to 0.00001.

\subsection{Evaluation}
We use the sed\_eval toolbox provided by the challenge~\cite{SED_EVAL} to evaluate the predictions made by R-CRNN.
Event-based error rate (ER) and event-based F1 score are the metrics used for evaluating the performance of AED methods in DCASE 2017.
In the challenge, these metrics are calculated using onset only condition with a collar of 500 ms. 

\subsection{Results}
\begin{table}[t]
\begin{center}
    \begin{tabular}{| l | l | l | l| l|}
    \hline
     Method & Babycry & Glass. & Gunshot & Avg. \\ \hline
     CNN & 0.31 & 0.13 & 0.21& 0.22\\ \hline
     CRNN\_GRU& 0.28 & 0.18 & 0.18 & 0.21 \\ \hline
     CRNN\_BiGRU & 0.14 & 0.08& 0.15& 0.12\\ \hline
    \end{tabular}
\end{center}
\caption{Event-based error rate of using different types of feature extractor for R-CRNN on DCASE 2017 development set.}\label{table:feature_ex}
\end{table}

\textbf{Experiments on Different Feature Extractors.}
To investigate the performance of different feature extractors, we experimented with different architectures to generate the high-level feature map for the region-based network.
Three different types have been tested: convolutional network (CNN), CRNN with an uni-directional GRU layer (CRNN\_GRU), and CRNN with a bi-directional GRU layer (CRNN\_BiGRU).
We set $M$ to 512 and use 100 units in the GRU layer.
For CNN, we use the same architecture as shown in Figure~\ref{CRNN} without the Bi-GRU layer.
For CRNN\_GRU, we replace the Bi-GRU layer in Figure~\ref{CRNN} with a uni-directional GRU layer.
Table~\ref{table:feature_ex} shows the results on DCASE 2017 development set.
Note that all of these results used pre-trained weights from the binary classification task as the weight initialization.
By adding a bi-directional GRU layer on top of the convolutional network, the generated high-level feature map contains long-term temporal contextual information and ER is reduced from 0.22 to 0.12.
Based on these results, we chose CRNN\_BiGRU as the feature extractor through our experiments.

\begin{table}[t]
\begin{center}
    \begin{tabular}{| l | l | l | l|}
    \hline
     Hyper-parameters & Babycry & Glassbreak & Gunshot \\ \hline
     \# units in Bi-GRU & 50 & 50 & 100\\\hline
     M & 128 & 256 & 256\\\hline
     \# parameters & 1,110,874& 1,250,394& 1,782,094\\\hline
    \end{tabular}
\end{center}
\caption{R-CRNN hyper-parameters for each target class.}\label{table:hyper_para}
\end{table}

\noindent
\textbf{Hyper-parameter Search.}
After selecting the feature extractor, we did a hyper-parameter search for R-CRNN.
The grid search covers on the number of units in Bi-GRU layer: \{50, 100\}; and the number of nodes in the dense layers in the final classifier (M): \{64, 128, 256, 512\}.
The best performing R-CRNN hyper-parameters and the number of parameters in the model for each target class are listed in Table~\ref{table:hyper_para}.

\begin{table}[t]
\begin{center}
    \begin{tabular}{| l | l | l | l| l|}
    \hline
     Method & Babycry & Glass. & Gunshot & Avg. \\ \hline
     W/o pre-training & 0.23 & 0.17 & 0.37 & 0.26 \\ \hline
     Pre-training & 0.25 & 0.09 & 0.35 & 0.23 \\ 
     (fixed CRNN) & & & & \\ \hline
     Pre-training & 0.09 & 0.04& 0.14& 0.09\\ 
     (finetune CRNN) & & & & \\ \hline
    \end{tabular}
\end{center}
\caption{Event-based error rate of using different pre-training methods for R-CRNN on DCASE 2017 development set.}\label{table:comp_pretraining}
\end{table}

\noindent
\textbf{Experiment on Different Pre-training Settings.}
To investigate the effect of pre-training on the feature extractor CRNN, we tested three different settings for pre-training.
We use the best performing hyper-parameters shown in Table~\ref{table:hyper_para} to build R-CRNN, and experiment on three settings:
1) \textit{Without pre-training:} R-CRNN is trained from scratch.
2) \textit{Pre-training (fixed CRNN):} We use the pre-trained weights of CRNN for weight initialization, and treat CRNN as a fixed feature extractor. There is no updates in CRNN during the training of R-CRNN.
3) \textit{Pre-training:} We use the pre-trained weights of CRNN for weight initialization, and all parameters are trainable during the training of R-CRNN.
Table~\ref{table:comp_pretraining} shows the results of using different pre-training methods.
Using the pre-trained CRNN as a fixed feature extractor works slightly better than without any pre-training.
By making all parameters trainable in the training of R-CRNN, the performance improves significantly in all target events.
We suspect that fixing the weights of CRNN constraints the search in the weight space during gradient descent in the training of R-CRNN, which decreases the performance.


\begin{table}[t]
\begin{center}
    \begin{tabular}{ l | l | l | l |l |}
     \cline{2-5}
     & \multicolumn{2}{|c|}{Development}& \multicolumn{2}{|c|}{Evaluation}\\
     \hline
     \multicolumn{1}{|l|}{Method} & ER & F-score & ER & F-score\\ \hline
     \multicolumn{1}{|l|}{R-CRNN} & 0.09 & 95.5\% & 0.23 & 87.9\% \\ \hline
     \multicolumn{1}{|l|}{1D-CRNN~\cite{Lim2017}}  & 0.07 & 96.3\% & 0.13 & 93.1\%\\
     \multicolumn{1}{|l|}{(Ranked 1st)} &&&& \\\hline
     \multicolumn{1}{|l|}{CRNN~\cite{Cakir2017}} & 0.14 & 92.9\% & 0.17 & 91.0\%\\
     \multicolumn{1}{|l|}{(Ranked 2nd)} &&&&\\\hline
     \multicolumn{1}{|l|}{DNN/CNN~\cite{Phan2017}} & 0.19 & 89.8\% & 0.28 & 85.3\%\\
     \multicolumn{1}{|l|}{(Ranked 3rd)} &&&&\\\hline
     \multicolumn{1}{|l|}{R-FCN~\cite{Kaiwu2017}} & 0.18 & 90.5\% & 0.32 & 82.0\%\\\hline
     \multicolumn{1}{|l|}{Baseline} & 0.53 & 72.7\% & 0.64& 64.1\%\\\hline
    \end{tabular}
\end{center}
\caption{Performance of different methods on DCASE 2017 development set and evaluation set.}\label{table:comp_others}
\end{table}

\noindent
\textbf{Comparison with Other Methods.}
Table~\ref{table:comp_others} shows the results (event-based ER and F-score) of R-CRNN, and the comparison with other AED methods on DCASE 2017 development set and evaluation set.
On the development set, R-CRNN without using ensemble method achieves performance comparable to 1D-CRNN~\cite{Lim2017}, which uses ensembles consist of up to 5 models for each target event.
Compared to the other region-based network for AED (R-FCN~\cite{Kaiwu2017}) with an ER of 0.18 on the development set, our method reduced the ER to half.
The major difference between R-CRNN and R-FCN is that we add a recurrent layer into R-CRNN to contain long-term temporal contextual information, where R-FCN is a fully convolutional network.
We thus observe that adding recurrent layers is advantageous for region-based networks on AED.
The results from experiments of different feature extractor also support this observation.

On the evaluation set, R-CRNN performs worse than the top 2 methods among all submissions in the challenge.
Note that both of the top 2 methods use ensemble method.
The ER increases from 0.09 on the development set to 0.23 on the evaluation set.
We suspect that R-CRNN suffers from over-fitting due to the high model complexity of region-based networks and limited amount of training data.
The region-based network for visual detection used weights pre-trained on ImageNet~\cite{ImageNet}, which contains 1.43 million annotated samples.
Our model is trained on a set of synthetic data that contains about 100 real samples for each target event.
We believe that using a larger training set with more non-synthetic samples may alleviate this over-fitting issue.


\section{Conclusions}
A Region-based Convolutional Recurrent Neural Network is proposed for AED.
There is no post-processing needed for converting the prediction from frame level to event level, since R-CRNN can be trained on a multi-task loss function that optimizes the classification and localization of audio events simultaneously.
On DCASE 2017 development and evaluation sets, R-CRNN is the best performing method without using ensemble method.
On the development set, our method achieves performance comparable to the method ranked 1st in the challenge~\cite{Lim2017}, which used an ensemble up to 5 models for each target event.
To utilize the temporal information in audio signals, a recurrent layer is added into R-CRNN to contain long-term temporal context.
Our method performs significantly better than the other region-based method (R-FCN)~\cite{Kaiwu2017}, which is a fully convolutional network.


\bibliographystyle{IEEEtran}

\bibliography{mybib}

\end{document}